\documentclass[12pt,a4paper,DIV=12]{scrartcl}
\pdfoutput=1
\usepackage[utf8]{inputenc}
\usepackage[T1]{fontenc}
\usepackage{microtype,textcomp}
\usepackage{graphicx,subfig,color}
\usepackage{enumitem}

\usepackage[format=plain,labelfont={sf,bf},textfont=sf]{caption}

\usepackage{amsmath,amsfonts,amssymb,isomath,bm}
\DeclareMathOperator{\sech}{sech}
	
\usepackage{ifthen}
\makeatletter
\newcommand{\greeksym}[1]{\usefont{U}{psy}{m}{n}#1}

\newcommand{\allmodesymb}[2]{\relax\ifmmode{\mathchoice
{\mbox{\fontsize{8.9}{\tf@size}#1{#2}}}
{\mbox{\fontsize{8.9}{\tf@size}#1{#2}}}
{\mbox{\fontsize{6.4}{\sf@size}#1{#2}}}
{\mbox{\fontsize{5.2}{\ssf@size}#1{#2}}}}
\else
\mbox{#1{#2}}\fi}
\makeatother

\newcommand{\ubeta}{\allmodesymb{\greeksym}{b}}

\newcommand{\mail}[1]{\href{mailto:#1}{\texttt{#1}}}

\usepackage{abstract}
\newcommand{\beginsupplement}{
        \setcounter{table}{0}
        \renewcommand{\thetable}{S\arabic{table}}
        \setcounter{figure}{0}
        \renewcommand{\thefigure}{S\arabic{figure}}
     }

\usepackage[affil-sl]{authblk}
	
	\author[1]{Lorenzo Pattelli\thanks{Corresponding author}}
	\author[1]{Romolo Savo\thanks{Current affiliation: \textsl{Laboratoire Kastler Brossel, UMR 8552, CNRS, Ecole Normale Sup\'{e}rieure, Universit\'{e}
Pierre et Marie Curie, Coll\`{e}ge de France, 24 rue Lhomond, 75005 Paris, France}}}
	\author[2]{Matteo Burresi\thanks{Current position: \textsl{Gestione SILO Srl, via di Castelpulci 14/d, 50010 Scandicci (FI), Italy}}}
	\author[1,3]{Diederik S. Wiersma}
	\affil[1]{European Laboratory for Non-linear Spectroscopy (LENS), Universit\`{a} di Firenze, 50019 Sesto Fiorentino (FI), Italy}
	\affil[2]{Istituto Nazionale di Ottica (CNR-INO), Largo Fermi 6, 50125 Firenze (FI), Italy}
	\affil[3]{Universit\`{a} di Firenze, Dipartimento di Fisica e Astronomia, 50019 Sesto Fiorentino (FI), Italy}

\usepackage{txfonts}
\usepackage[scaled=.86]{helvet}

\usepackage[version=3]{mhchem}

\usepackage[range-units = single,range-phrase = \text{--},separate-uncertainty = true,detect-all]{siunitx}
	\DeclareSIUnit\linepair{lp}
	\DeclareSIUnit\pixels{px}

\newcommand{\mytitle}{Spatio-temporal visualization of light transport in complex photonic structures}
\title{\mytitle}
\date{}

\usepackage[style=nature,backend=bibtexu,sorting=none]{biblatex}	
	
	\bibliography{references}

\usepackage[bookmarks,hidelinks,pdfauthor={Pattelli et al.},pdftitle={\mytitle}]{hyperref}

\begin{document}

\maketitle
\vspace{-1.2cm}

\begin{abstract}
Spatio-temporal imaging of light propagation is very important in photonics because it provides the most direct tool available to study the interaction between light and its host environment. Sub-ps time resolution is needed to investigate the fine and complex structural features that characterize disordered and heterogeneous structures, which are responsible for a rich array of transport physics that have not yet been fully explored. A newly developed wide-field imaging system enables us to present a spatio-temporal study on light transport in various disordered media, revealing properties that could not be properly assessed using standard techniques. By extending our investigation to an almost transparent membrane, a configuration that has been difficult to characterize until now, we unveil the peculiar physics exhibited by such thin scattering systems with transport features that go beyond mainstream diffusion modeling, despite the occurrence of multiple scattering.
\end{abstract}

\begin{refsection}

\section*{Introduction}
Light represents a useful and versatile probe to study our world. By tracking its evolution in the domains of time, space, frequency and reciprocal space, we are able to extract a wealth of information regarding the surrounding environment. Further insight is often gained when we are technically able to track this evolution in more than one such domain simultaneously, as demonstrated by a number of recent results from diverse research fields\autocite{drexler2001ultrahigh, obrig2003beyond, dunsby2003techniques, engelen2007ultrafast, minardi2010three, bassi2010time, velten2012recovering, eggebrecht2014mapping}. Of the different approaches, spatio-temporally resolved techniques undoubtedly offer the most straightforward approach, as underlined by the continuously growing interest in fast photography\autocite{gao2014single, nakagawa2014sequentially, li2014single} and imaging\autocite{gundlach2008femtosecond, sperling2012direct, farid2013dynamics, grumstrup2015pump} applications.
Further development of similar techniques is key to our ability to investigate increasingly more complex media, such as disordered or heterogeneous media, which are ubiquitous in most fields of study, from atmospheric physics to biology and cultural heritage preservation. In this context, one main task of optical investigations is to characterize the light transport process determined by multiple scattering. This characterization is typically achieved through repetitive measurements to retrieve meaningful microscopic transport parameters, such as the transport mean free path $l_\text{t}$ and the absorption length $l_\text{a}$, which are directly related to the properties of the investigated medium (that is, the internal structure, composition and/or density), and are key parameters to consider from both the fundamental and the application points of view.

Here, we experimentally combine wide-field spatial imaging and sub-picosecond (ps) time resolution and show how the acquired multi-dimensional dataset allows us to characterize previously inaccessible scattering systems and to improve the accuracy of measurements in standard situations. The ultrafast time resolution needed for such applications represents a major experimental challenge. Established fast imaging techniques involve electronic-based devices such as streak cameras or sequential scanning over a given sample in a pump-probe configuration. Despite providing very accurate results, many of these techniques are limited by low-time resolutions\autocite{sperling2012direct} or a long acquisition time\autocite{grumstrup2015pump}. Conversely, wide-field images at sub-ps time scales are easily obtained with optical gating techniques; however, their limited accuracy has limited their use to qualitative pattern/shadowgram recognition rather than to absolute intensity measurements across the entire field of view. In other words, optical gating has been previously exploited as a convenient method to see through a turbid medium obliterating the complex light transport arising from its structure, not to study it directly.

In this study, we investigate light transport focusing on three classes of disordered media that are extremely relevant for practical applications and yet still very difficult to characterize accurately, namely samples with limited thickness, large-scale heterogeneities and semi-transparent membranes. In each of these cases, transport properties are retrieved by exploiting, for the first time, an all-optical gating setup, which we developed to offer both wide-field acquisition and high quantitative accuracy on a sub-ps time scale. Our spatio-temporal investigation allows us to grasp the evolution of light transport in its entirety, consequently revealing effects that were either overlooked or misinterpreted to date, including a peculiar regime occurring in ultra-thin disordered media that becomes apparent only when the spatial and temporal domains are studied simultaneously and would not be detectable with other state-of-the-art techniques. By comparing the outcome of our analysis with that of the mainstream evaluation model based on the diffusive approximation (DA), we describe how single-domain investigations performed to date are susceptible to particularly deceptive artifacts that can be clearly appreciated in a multi-domain dataset. At the same time, as widely suggested in the literature, we discuss how much more robust interpretations, even within the diffusive framework, can be given in terms of transverse transport, to which we gain direct access through our technique.

\section*{Materials and Methods}
\begin{figure}
\centering
\includegraphics[width=\textwidth]{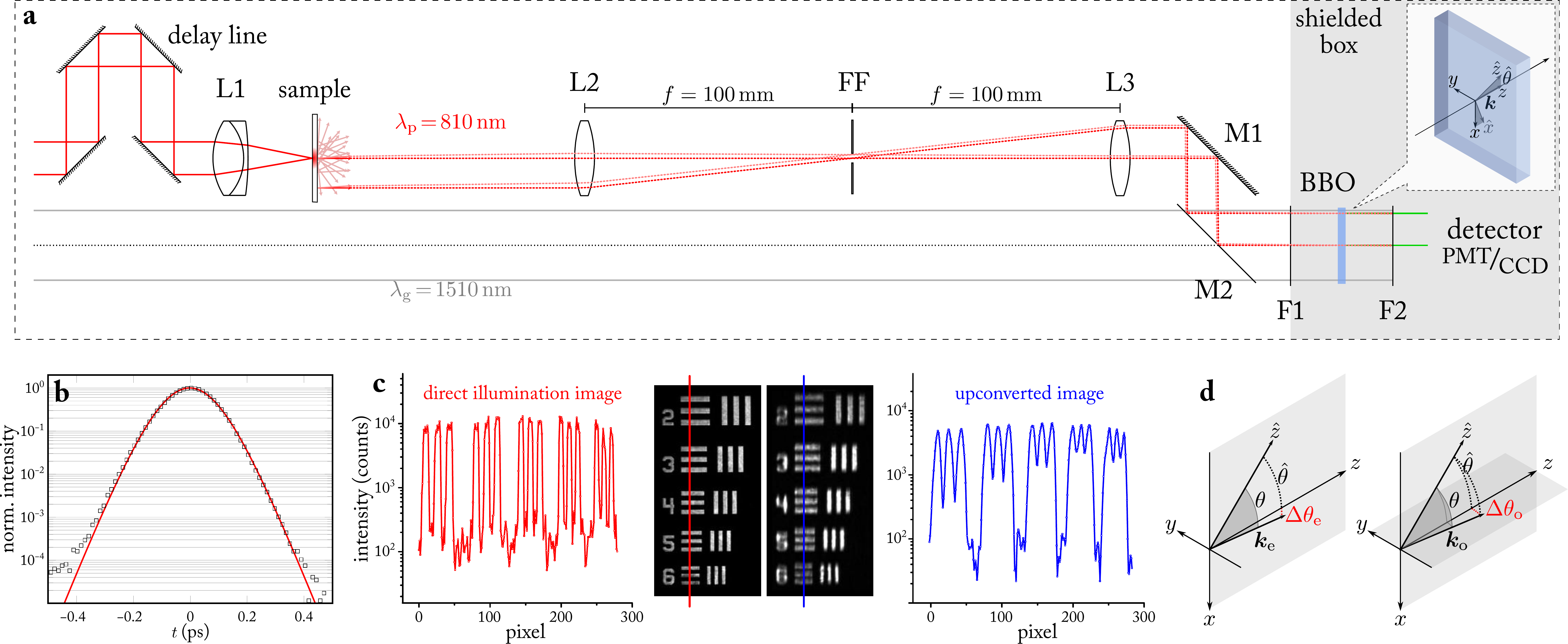}
\caption{\textbf{Experimental setup and spatio-temporal resolution.} \textbf{a} Two linearly polarized \si{\femto\second}-pulses are provided by a Ti:Sa laser (probe in red) and a synchronously pumped optical parametric oscillator (gate in gray). The probe beam interacts with a slab sample, which is imaged on a BBO crystal with a 4$f$ lens configuration. A small aperture (FF) allows the $k$ vectors to be isotropically filtered, while a dichroic mirror (M2) guarantees signal and gate spatial superposition. A long-pass (F1) and a bandpass filter (F2) block the visible and unconverted light, respectively. \textbf{b} A typical cross-correlation time-resolved measurement acquired with a PMT. A fit with the cross-convolution of two $\sech ^2$ pulses returns a FWHM of \SI{174}{\femto\second}. \textbf{c} The spatial information carried by the pulse is retrieved using a CCD camera, whose resolution is influenced by the sum-frequency process. \textbf{d} A geometric sketch qualitatively showing how the phase-matching condition results in different angular acceptances in the extraordinary and ordinary directions. A slightly misaligned $\bm{k}_\text{e}$ and a largely misaligned $\bm{k}_\text{o}$ might give the same $\mathrm{\Delta}\theta$, therefore explaining the sharper resolution in the $y$ direction.}
\label{fig:setup}
\end{figure}

\paragraph{Experimental setup} The optical scheme employed to resolve the time and space propagation of light in complex media is based on the cross-correlation gating technique\autocite{shah1988ultrafast, yodh1990pulsed}, shown schematically shown in Fig.~\ref{fig:setup}a.
Two synchronous, collinear probe and gate pulses at different wavelengths impinge on a \SI{2}{\milli\meter} thick $\ubeta$-Barium borate (BBO) crystal. When the two beams overlap both spatially and temporally, a sum-frequency signal is generated with intensity proportional to the cross-correlation of the original pulses as a function of the delay-line position. This technique has been used to fully characterize ultrafast pulse propagation\autocite{trebino2000frequency, savo2014walk, burresi2014bright}, generally disregarding the spatial distribution of the converted signal (cfr.\ Fig.~\ref{fig:setup}b).
Nevertheless, cross-correlation gating techniques offer several advantages, such as being unaffected by optical chirp (in contrast to interferometric methods) and being compatible with collinear schemes because possible spurious second harmonic contributions from gate/probe beams, if any, can still be spectrally separated. Indeed, a collinear geometry is particularly convenient to mitigate possible distortion of upconverted images passing through the BBO, which is a crucial aspect for accurate spatially resolved experiments\autocite{potenza2004three}. Finally, the gate and probe beams are perfectly exchangeable, therefore allowing us to easily investigate any sample at both wavelengths.

\paragraph{Upconversion imaging} To study the dynamics of light emerging from a complex specimen, we image its exit surface on the BBO crystal. There, the original image at the probe wavelength is upconverted to a different (signal) wavelength by sum-frequency generation with a gate pulse. Non-linear upconversion is a polarization- and wavevector-sensitive process, and both of these aspects must be taken into account when designing an imaging apparatus. In the current study, which addresses multiple scattering systems, the polarization sensitivity is not relevant because all polarization channels have the same optical properties. However, the setup inherently allows to investigate separately different polarization channels. In the phase-matching condition, the non-linear crystal converts only those wavevectors laying within a relatively narrow paraxial range in reciprocal space, which defines the angular acceptance of the BBO crystal (see SI). This acceptance angle is weighted differently in the $x$- and $y$-directions of the BBO surface (that is, along the image plane, see Figs.~\ref{fig:setup}c and d). A double telecentric apparatus in a $4f$ configuration ensures that the paraxial component emerging from the sample is preserved at the crystal interface. This allows us to obtain a quantitatively accurate reproduction of the original intensity distribution at each time step for each point of the object field. This accuracy condition holds as long as the transmitted wavevector distribution, however complex, does not depend on the point and time of exit, as is the case with multiple scattered light in disordered samples. In this case, the intensity carried by any (fixed) subset of wavevectors can be considered to be strictly proportional to the total transmitted intensity.

As seen in Figure~\ref{fig:setup}c, the reduced set of available wavevectors degrades the sharpness of the retrieved image differently along the two axes. To obtain the same resolving power in both the $x$ and $y$ directions, the aperture of a Fourier filter (FF) can be reduced to act as a tunable isotropic wavevector filter. Several alternative upconverting configurations have been reported in the literature regarding full-frame imaging applications\autocite{firester1970image, chiou1971geometric, faris1994upconverting, devaux1995ultrahigh, abraham2000real, lantz2008parametric, pedersen2009enhanced, bassi2010time, dam2010high}, most of which employ a non-linear crystal in the Fourier plane of the imaging optics rather than in the image plane. Both configurations present different advantages, and the preference for their adoption depends on the specific application. In the common Fourier plane configuration, angular acceptance limits upconversion uniformity over the entire field of view; in our case, the image is formed with a cut-off set of $k$ vectors (thus with lower resolution) while guaranteeing spatially uniform upconversion efficiency. Therefore, we adopted the latter configuration because our diffuse signal typically does not exhibit any sharp features, and a spatially uniform response represents a critical condition for attaining our quantitative accuracy goal over multiple orders of magnitude. The generated signal is collected by a lens inside a shielded box where a flip mirror allows us to switch between different detection schemes.

We initially performed standard time-resolved measurements using a photomultiplier tube (PMT), that is, collecting intensity without any spatial resolution. Figure~\ref{fig:setup}b shows a typical cross-correlation measurement of our probe pulse passing through a test target together with a fit obtained from the convolution of two squared-hyperbolic secant functions (see SI). . The measurement does not exhibit any relevant satellites over several orders of magnitude. By switching the photon counter with a CCD camera, a transient image is detected (Fig.~\ref{fig:setup}c) that shows the spatial resolution of our setup ($\sim \SI{22.6}{\linepair\per\milli\meter}$ in $x$ and $\sim \SI{11.3}{\linepair\per\milli\meter}$ in $y$ at \num{0.5} contrast), which is comparable to previous publications\autocite{faris1994upconverting, bassi2010time, dam2010high, abraham2000real}. The final resolution is obtained by reducing the aperture of the FF to $<\SI{1}{\milli\meter}$, until the same (isotropic) resolution of $\sim \SI{11.3}{\linepair\per\milli\meter}$ is obtained in both directions.

With regards to the overall sensitivity of the upconversion process, previous measurements performed with the same setup\autocite{burresi2014bright} give an estimate of the signal attenuation at which the noise level is eventually reached. Starting from a typical pulse energy of $\sim \SI{1}{\nano\joule}$ at a probe wavelength of \SI{810}{\nano\meter}, an upconverted signal can still be detected after a damping of 8 decades, taking into account the fraction of the probe beam that is usually lost by diffuse reflection and the limited solid angle subtended by the collection optics. In addition, several other parameters can be adjusted to further enhance sensitivity, such as increasing the integration time of the detector, using a non-linear crystal with a higher upconversion efficiency and a larger angular acceptance (for example, bismuth triborate) or increasing the fluence on the crystal. To this purpose, it is particularly convenient to increase the gate beam intensity as much as possible to avoid any alteration of the sample.

\paragraph{Data evaluation models}
By accessing both the spatial and temporal evolution of light emerging from a complex environment, an incredibly rich picture emerges that needs to be analyzed with the proper evaluation tools. Exact modeling of energy transport in disordered samples is provided by the radiative transfer equation, which models light propagation as a random walk of energy packets undergoing scattering and absorption events. The characteristic length, $l_\text{s}$, over which a packet is scattered on average is called the scattering mean free path, whereas the anisotropic scattering factor, $g$, expresses the average degree of directionality ranging from completely random ($g=0$) to perfectly forward ($g=1$). Both these parameters are directly linked to the microscopic structure of the sample and to the properties of its basic scattering elements. However, a similarity relation exists\autocite{graaff1993similarity, van2012multiple} equating a random walk with a certain $l_\text{s}$ and $g$ to a completely isotropic random walk with a rescaled mean free path of $l_\text{t} = l_\text{s}/(1-g)$.
. Thanks to this relationship, light transport is often studied in the isotropic diffusion approximation (DA) regime, where a single diffusion coefficient, $D = l_\text{t} v/3$, is used as the only transport parameter, with a consequent loss of insight concerning the microscopic properties of the sample. Moreover, the convenient description offered by the DA is known to be valid only when the transport mean free path is much smaller than the sample size. Conversely, its applicability to optically thin media is often debated and a standard approach to measure transport parameters in this regime is still missing. Several attempts have been made to derive extensions of the diffusive model for optically thin media\autocite{elaloufi2002time, xu2002photon, garofalakis2004characterization} , and alternative techniques to measure transport parameters have been proposed that avoid using diffusion theory in the first place\autocite{hammer1995optical, kienle1996determination, pifferi1998real, alerstam2008white, leonetti2011measurement, nilsson1995measurements, dam2005real, karlsson2012inverse, svensson2013exploiting} and are based on single-domain (for example, temporal, spatial or angular) investigations.

In this study, we compare analyses of our experimental data with the mainstream diffusion theory and a Monte Carlo model of light transport. While the former method retains its appeal due to its simplicity, the latter represents a direct implementation of the radiative transfer theory, providing a computationally expensive but asymptotically exact solution to the transport problem. This allows us to unveil some of the particularly deceptive ways that the DA can fail when dealing with certain types of media.

\section*{Results and Discussion}
In the following section, we present a few relevant experimental cases where exploiting our newly developed ultrafast imaging (UFI) setup, we were able to reveal structural and optical features of the investigated samples that could not have been probed unambiguously with current state-of-the-art time-resolved techniques and data evaluation models. These claims are further validated and quantified by a direct comparison of the experimental data with Monte Carlo simulations. The main advantages offered by our experimental technique, as well as other possible applications, are discussed in the final subsection.

\paragraph{Limited thickness artifacts} We first tested our setup with the simplest possible disordered slab, a homogeneous isotropic sample made of \ce{TiO2} nanoparticles embedded in a polymer matrix (see SI). The sample thickness was \SI{203}{\micro\meter} with an average refractive index at \SI{810}{\nano\meter} of \num{1.52}.

\begin{figure}
\centering
\includegraphics[width=\textwidth]{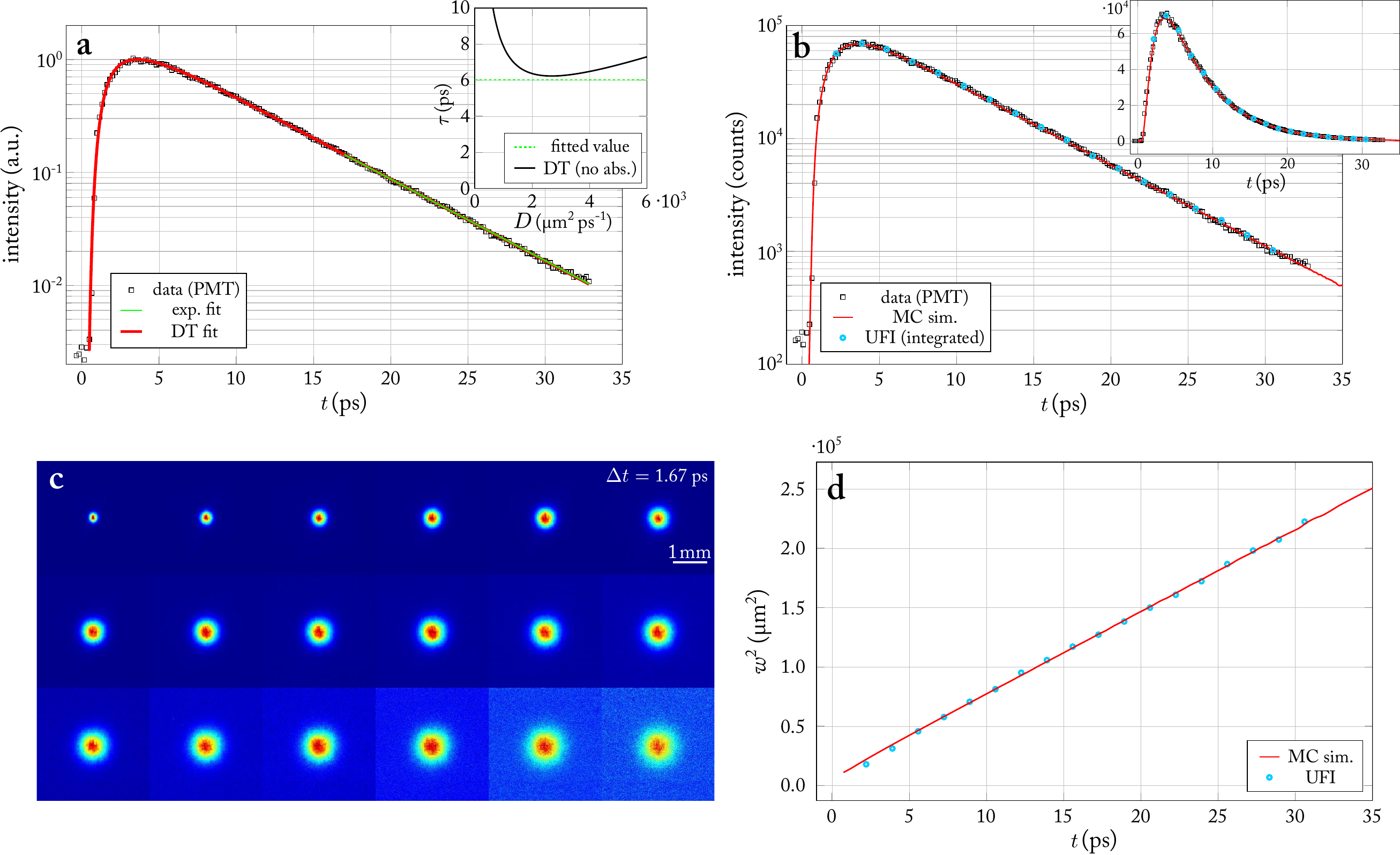}
\caption{\textbf{Homogeneous sample data analysis.} \textbf{a} Experimental data (black squares) acquired with the photon counter (PMT) compared to an exponential fit (solid green line), and a diffusion theory fit with $D$ and $\mu_\text{a}$ as free parameters (solid red line). The inset shows the asymptotic decay time expected according to the DA for a sample with $L = \SI{203}{\micro\meter}$, $n=\num{1.52}$ and $\mu_\text{a} = \SI{0}{\per\micro\meter}$ (solid black line). The experimentally retrieved value (dashed green line) is incompatible with the DA unless some absorption is introduced. \textbf{b} A Monte Carlo (MC) solution (solid red line) of the radiative transfer equation showing perfect agreement with the experimental data for a non-absorbing sample. The integrated intensity from UFI frames and the PMT data (cyan circles) are shown. \textbf{c} A set of frames acquired at different delays with our UFI setup. Each frame is averaged over different disorder realizations and is displayed normalized to its own maximum intensity. \textbf{d} Comparison of the experimental and simulated mean square width.}
\label{fig:reference}
\end{figure}

Figure~\ref{fig:reference}a shows a normalized time-resolved transmission profile, in addition to a fit using the DA relation, which shows excellent agreement with the experimental data. The fit was performed with the diffusion coefficient, $D$, and absorption coefficient, $\mu_\text{a} = l_\text{a}^{-1}$, as the only free parameters, assuming $D = l_\text{t}v/3$ and $v=c/n$.
The fit returns $l_\text{t} = \SI{24.1}{\micro\meter}$ and $l_\text{a} = \SI{12.0}{\milli\meter}$ (albedo \num{0.998}).
The ratio between the thickness of the slab and its transport mean free path (referred to as the optical thickness) is commonly used to express the amount of multiple scattering in a sample. According to the fit, we obtained an optical thickness of $\sim \num{8.4}$; a value of 8 is usually considered to be the rule-of-thumb lower limit above which diffusive transport is regarded as an acceptable approximation\autocite{elaloufi2004diffusive}. Nonetheless, the retrieved coefficients predicted a total absorption of $A \sim \SI{8}{\percent}$which is much higher than expected because both the host polymer and the \ce{TiO2} nanoparticles are known to have vanishing absorption at near infrared wavelengths. Even considering a non-absorbing sample with an equal thickness and refractive index, the lowest possible decay time that diffusion theory could predict ($\tau_\text{min}^\text{DA} = \SI{6.23}{\pico\second}$, see Fig.~\ref{fig:reference}a, inset) would still be appreciably longer than that determined experimentally ($\tau = \SI{6.01+-0.08}{\pico\second}$). The DA accommodates for such discrepancies by introducing an absorption artifact. This can intuitively be seen in its relationship for the asymptotic decay time, $\tau^{-1} = \pi^2 D/L_\text{eff}^2 + \mu_\text{a}v$ (where $L_\text{eff}$ is the effective thickness depending on the refractive index boundary conditions), the value of which can be made arbitrarily small by assuming the presence of absorption.

In this case, the analysis of the time evolution of the mean square width of the transmitted profile is particularly interesting in that it is inherently free from this absorption-to-scattering crosstalk effect\autocite{cherroret2010transverse, hu2008localization, sperling2012direct}. The mean square width is defined as the variance, $w^2 (t) = \int r^2 I(\bm{r},t)\, \mathrm{d}^2 \bm{r} \big/ \int I(\bm{r},t)\, \mathrm{d}^2 \bm{r}$, where $\bm{r}$ is a two-dimensional vector in the transmission plane, and can be determined if the full spatio-temporal evolution of the profile $I(\bm{r},t)$ is known. Due to its normalization, any amplitude factor (such as absorption) applied to the profile will cancel out exactly at any time, leaving the mean square width unaffected. We therefore further investigated the sample using our imaging configuration (which could in principle also substitute the photon counter following frame integration, see Fig.~\ref{fig:reference}b). The mean square width, as extracted by the obtained frames, clearly shows a linear increase (Figures \ref{fig:reference}c-d) that can easily be interpreted within the DA, which predicts $w^2 (t) = 4Dt$. The experimental slope, as obtained from a linear fit, corresponds to $D_\text{exp}^\text{fit} = \SI{1746}{\micro\meter\squared\per\pico\second} \rightarrow l_\text{t} = \SI{26.6}{\micro\meter}$, which is appreciably larger than the value retrieved from the time-resolved curve.
We therefore ran Monte Carlo simulations (see SI) to resolve this discrepancy, and eventually succeeded in simultaneously reproducing both the experimental time-resolved and mean square width evolution datasets using a single transport parameter, $l_\text{t}$ (Figures \ref{fig:reference}b and d). The Monte Carlo inversion procedure gives a value of $l_\text{t} = \SI{25.5}{\micro\meter}$ (again corresponding to an optical thickness of $\sim \num{8}$) which is perfectly compatible with both datasets without introducing absorption.

It should be noted that the transport mean free path, as retrieved from the mean square width slope in the diffusive framework, results in just a \SI{4}{\percent} overestimation of the actual best value retrieved by the Monte Carlo simulations. Such a level of accuracy is acceptable for a wide number of applications, where UFI techniques could be exploited to extend the applicability of the DA in this intermediate thickness range. Moreover, as widely suggested in the literature\autocite{cherroret2010transverse, hu2008localization}, estimating transport properties from the mean square width rather than from time-resolved data is much more accurate and straightforward because in the former case, \emph{a priori} knowledge of the absorption coefficient, refractive index contrast and sample thickness is not required. Indeed, the simple $4Dt$ dependence does not involve these parameters or the extrapolated boundary conditions. This simple determination of the slope is also insensitive to the exact determination of the time axis origin and the exact time-window considered for the fitting, both of which represent long-standing issues in the evaluation of time-resolved data\autocite{alerstam2008improved, bouchard2010reference}. In addition, for the same reason that the mean square width is independent of the absorption, it is also effectively independent of the integration time, as well as of any laser power fluctuations. Interestingly, the time-resolved DA can deceptively compensate for its limited accuracy by introducing an absorption artifact, but lacks a solution when the correct value of absorption is considered. This point is particularly relevant for investigations of optically thin specimens, such as biological tissues\autocite{tuchin2007tissue}.

As previously mentioned, our UFI technique is capable of replacing completely integrated measurements with comparable sensitivity (Fig.\ \ref{fig:reference}b). Exploiting the full spatio-temporal information, the resulting set of frames can be decomposed into a mean square width and a time-resolved curve, providing two almost independent datasets in a single measurement session. Indeed, the former does not depend on the integrated intensity of the individual frames, whereas the latter is primarily linked to the transport along the $z$ axis. In this respect, our perfectly matching Monte Carlo validation, which is conducted independently on two experimental datasets that depend differently on different sets of parameters, is a clear indication of the quantitative accuracy and reliability of our setup over many orders of magnitude in both the temporal and spatial domains.

\begin{figure}
\centering
\includegraphics[width=\textwidth]{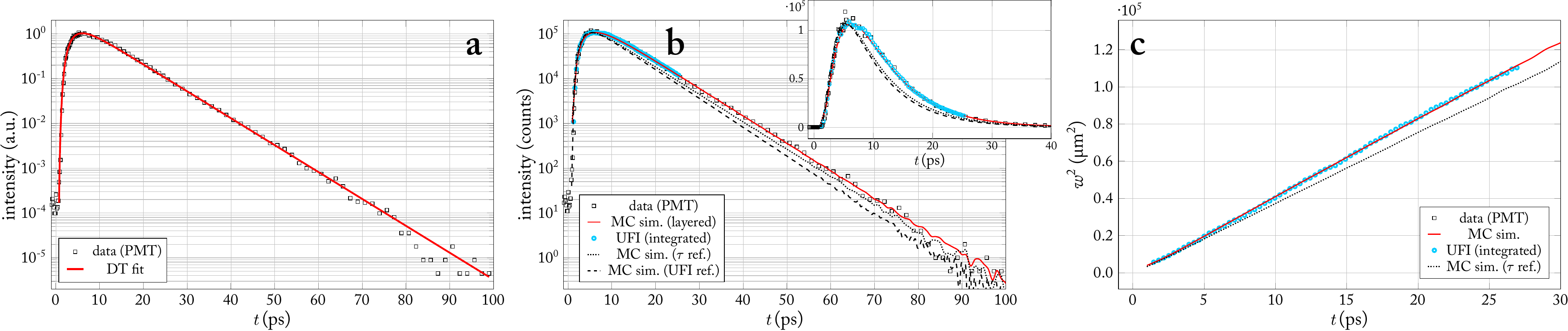}
\caption{\textbf{Heterogeneous sample data analysis.} \textbf{a} Photon counter (PMT) data (black squares) and best diffusive (DA) fit (solid red line) giving $l_\text{t} = \SI{12}{\micro\meter}$ and $l_\text{a} = \SI{13.2}{\milli\meter}$. \textbf{b}, \textbf{c} Comparison of the PMT and UFI data to a multi-layered simulation (solid red line) and two homogeneous simulations. The black-dotted line refers to the simulation that reproduced the experimental decay time ($l_\text{t} = \SI{14.2}{\micro\meter}$), while the black-dashed line shows the simulation that reproduced the mean square width slope ($l_\text{t} = \SI{15.7}{\micro\meter}$). The decay-time-reproducing simulation fails to reproduce the mean square width slope, while the simulation that shows perfect agreement with the mean square width slope poorly matches the time-resolved decay, therefore revealing the inconsistency of the homogeneity hypothesis. The Monte Carlo (MC) curves were rescaled to match the experimental peak count values.}
\label{fig:multilayer_MC}
\end{figure}

\paragraph{Large-scale heterogeneity artifacts}
We also used our UFI technique to investigate more diverse and complex media. A new polymer slab ($L = \SI{190}{\micro\meter}$) with a roughly doubled scatterer density was manufactured to obtain an optically thicker sample. Again, a perfect fit could be obtained in the diffusive framework (Fig.~\ref{fig:multilayer_MC}a) if the presence of absorption was included at an unlikely integrated value for the total absorption, $A \sim \SI{7}{\percent}$ ($l_\text{t} = \SI{12}{\micro\meter}$, $l_\text{a} = \SI{13.2}{\milli\meter}$). In sharp contrast to the previous experiment, a non-absorbing Monte Carlo fit was unable to reproduce the time-resolved data, even when the observed experimental decay time was perfectly matched (Fig.~\ref{fig:multilayer_MC}b, dotted black line). Furthermore, the mean square width exhibited a steeper than expected slope ($D^\text{fit}_\text{exp} = \SI{1050}{\micro\meter\squared\per\pico\second} \rightarrow l_\text{t} = \SI{16}{\micro\meter}$) as if the diffusion process was enhanced along the in-plane directions, as shown in Figure~\ref{fig:multilayer_MC}c.

\begin{figure}
\centering
\includegraphics[width=\textwidth]{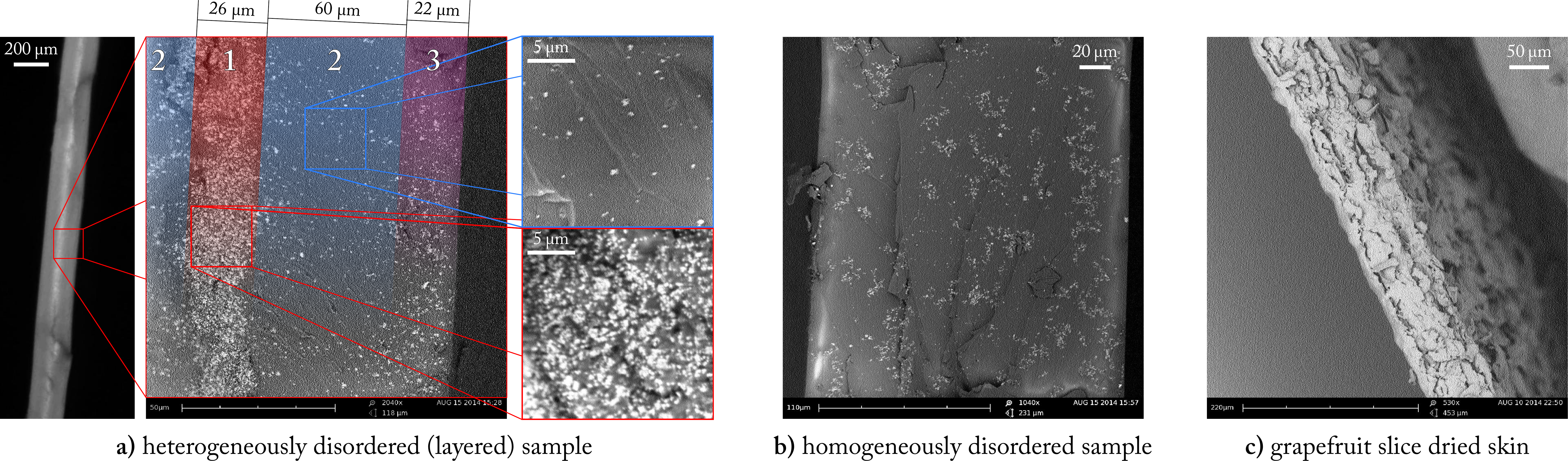}
\caption{\textbf{SEM side-view images of the analyzed samples.} \textbf{a} Heterogeneously disordered sample. The central inset shows a transverse section, which exhibits a layered symmetric structure that we modeled with a high-density central layer ($L_1 = \SI{26}{\micro\meter}$), a low density interstitial layer ($L_2 = \SI{60}{\micro\meter}$) and an intermediate density outer layer ($L_3 = \SI{22}{\micro\meter}$). The heterogeneity represented by the core layer is also appreciable at the optical microscope (left inset). The right insets highlight the density difference between regions 1 and 2. \textbf{b} A transverse section of the homogeneously disordered reference sample. Despite showing signatures of scatterer clustering, small-scale clusters/heterogeneities are homogeneously dispersed across the sample thickness and do not exhibit any layered arrangement. As confirmed by the excellent agreement obtained with a single parameter Monte Carlo simulation and other work on similar samples\autocite{vos2013broadband}, light perceives an effective, homogeneous scatterer density in contrast to the previous tightly layered sample. \textbf{c} Lateral section of the dried skin of the grapefruit slice.}
\label{fig:SEM}
\end{figure}

Driven by this discrepancy, we further investigated the sample with a scanning electron microscope (SEM), revealing a heavily layered modulation of the scatterer density compared to the more homogeneously dispersed one in our first sample (Figures \ref{fig:SEM}a and b).
Taking advantage of the versatility of Monte Carlo simulations, we modeled our sample after our SEM images as being composed of five layers with three different thicknesses and densities arranged symmetrically with respect to the central layer. A new Monte Carlo brute force fit assuming a fixed geometry and three free parameters, $l_\text{t1}$, $l_\text{t2}$, $l_\text{t3}$, was eventually able to perfectly reproduce both the time-resolved and mean square width data simultaneously, as shown in Figures~\ref{fig:multilayer_MC}b-c. We obtained transport mean free paths of $l_{t1} = \SI{3.5}{\micro\meter}$, $l_{t2} = \SI{21.5}{\micro\meter}$ and $l_{t3} = \SI{11}{\micro\meter}$, respectively, for the high, low and intermediate-density layers. The absorption coefficients were all set to $\mu_\text{a} = \SI{0}{\per\micro\meter}$.
In this case, the mean square width/time-resolved decomposition clearly allowed an unexpected degree of complexity to be revealed and characterized thanks to the rigid constraints of the double boundary datasets. Resolving this composite structure showed that a layered heterogeneity can, in principle, mimic the effects of anisotropic transport and therefore revealed a previously unaddressed physical effect. This is especially relevant given the pervasiveness of layered media (for example, coatings, atmospheric physics and biological tissues), which often exhibit counter-intuitive features that standard time-resolved techniques are still unable to explain, as in the case of light transport across the human forehead\autocite{comelli2007vivo}.

\begin{figure}
\centering
\includegraphics[width=0.8\textwidth]{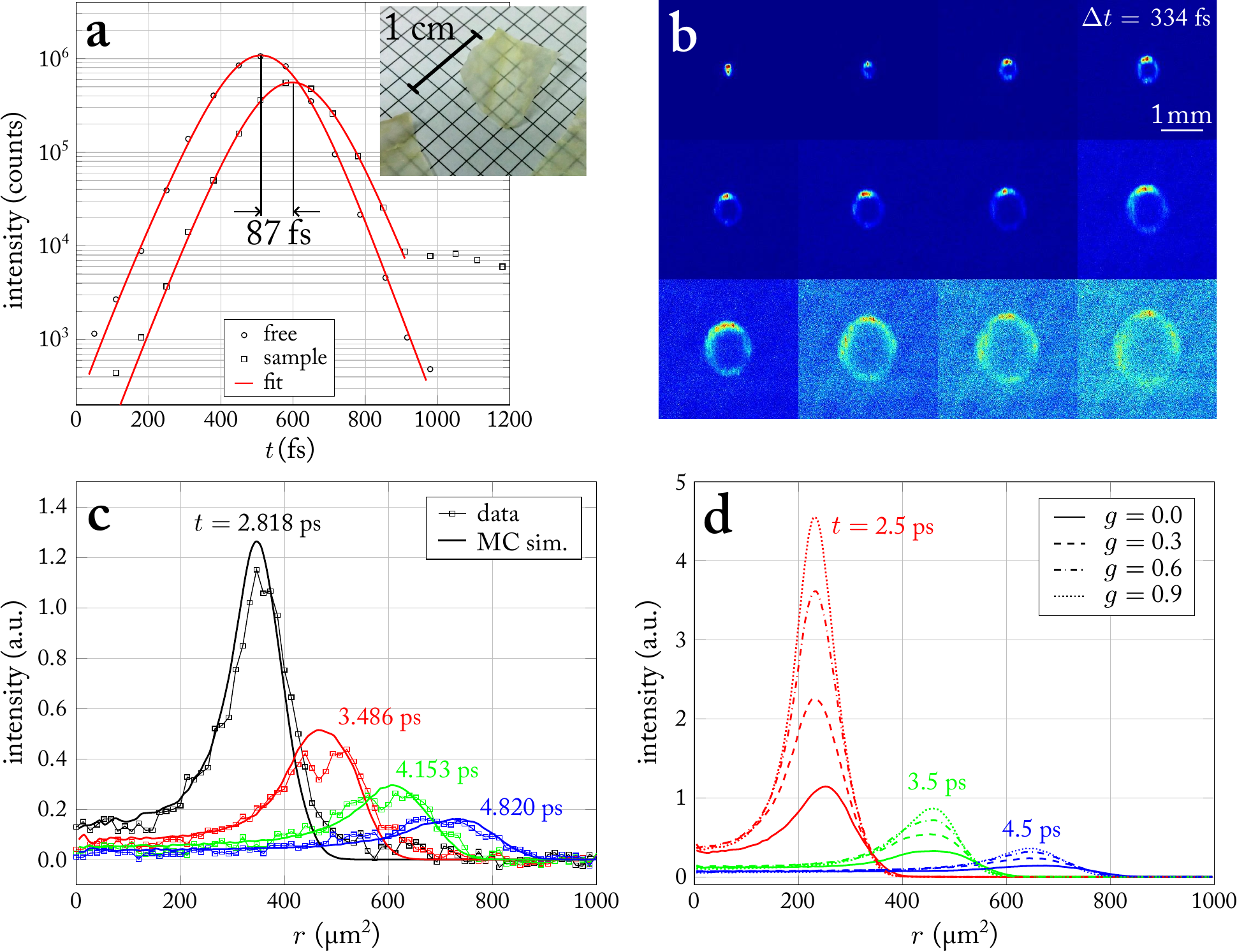}
\caption{\textbf{Grapefruit skin data analysis.} \textbf{a} The time shift induced by the sample in the probe path (black circles vs. black squares). Knowing its thickness from previous scanning electron microscopy images, the effective refractive index is found as $n = 1 + c\mathrm{\Delta}t/L \sim \num{1.31}$. The inset shows the actual appearance of the investigated specimens. \textbf{b} The frame set, each of which is averaged over a small portion of the sample. The transmitted intensity distribution shows a quasi-ballistic elliptic wavefront traveling at slightly different speeds (see SI). \textbf{c} Horizontal cross-cuts of the wavefront and a Monte Carlo fit with $g$ and $l_\text{s}$ as free parameters. The fit returns values of $g = \num{0.7}$ and $l_\text{s} = \SI{150}{\micro\meter}$. \textbf{d} Simulated radial intensity profiles in this extremely optically thin regime show a significant breakdown of the similarity relation.}
\label{fig:grapefruit}
\end{figure}

\paragraph{Ultrathin samples} Despite the advantages offered by decomposition into a mean square width and a time-resolved curve, it is clear that the raw, non-integrated data provide an irreducible set of information. A dramatic illustration of this point is obtained by studying a thin biological sample, such as a small strip cut from the dried skin of a slice of grapefruit. As the SEM image reveals (Figure~\ref{fig:SEM}c), its structure consists of a conglomerate of small flakes forming a corrugated slab $\sim \SI{85}{\micro\meter}$ thick. The dried sample appears to be a brittle, almost transparent membrane (Figure~\ref{fig:grapefruit}a, inset).

In this range of optical (and absolute) thicknesses, standard experimental techniques fall short because of the extreme time scales involved, causing common diffusive modeling to fail dramatically. Furthermore, most of the signal will be ballistically transmitted, carrying almost no information concerning the optical properties of the sample. As Figure \ref{fig:grapefruit}a shows, the light that is scattered inside the grapefruit membrane is 1/50th of the original intensity. Our ultrafast imaging technique allows us to selectively address the light that was retained for a longer time inside the sample whereas at the same time spatially inspecting it as it propagates along the main (transverse) slab extension.
Figure~\ref{fig:grapefruit}b shows a collection of frames acquired over a time-window of $\sim \SI{5}{\pico\second}$ after pulse injection, corresponding to a total path length greater than 13 times the sample thickness.
We see that the light spreads through the membrane with a well-defined wavefront traveling inside its main plane, resulting in a dramatic departure from any prediction compatible with the diffusive framework. However, standard time-resolved or steady state investigations would still just measure a single decay time and a bell-shaped profile, respectively, which could deceptively support an inappropriate interpretation in terms of the DA.

As we will show, our ultrafast time- and space-resolved experiment provides a deep insight into the light transport properties of complex systems, which requires the development of a novel analysis methodology. We demonstrate a proof-of-concept procedure to assess in-plane transport properties in this extremely optically thin regime. For the sake of simplicity, we will illustrate our investigation on a cross-cut of the profile assuming isotropic transport. The general treatment is analogous and will be the subject of future work.
Figure \ref{fig:grapefruit}c shows a comparison between experimental cross-cuts at different times and the corresponding time evolution of an isotropic Monte Carlo simulation that succeeds in reproducing the wavefront speed, intensity decay and overall shape at all given times. To achieve this collective agreement, the anisotropic scattering coefficient, $g$, needs to be considered as an additional parameter to express the degree of directional correlation between consecutive scattering events. As mentioned, diffusion theory is degenerate in $g$ as long as the transport mean free path $l_\text{t}$ is kept constant, as expressed by the similarity relation $l_\text{t} = l_\text{s}/(1-g)$. This is why retrieving any information concerning the angular scattering function is particularly challenging, and one must therefore resort to a regime where the DA ceases to hold\autocite{svensson2013exploiting}.
Indeed, as shown by Figure \ref{fig:grapefruit}d, the observed transverse intensity patterns exhibit a significant breakdown in this degeneracy condition in thin slabs, with $g$ playing a substantial role in determining the shape and evolution of the traveling wavefront, even when $l_\text{t}$ is constant. In particular, while the outer wavefront propagates almost ballistically (see SI), its instantaneous position and peak intensity vary appreciably with different combinations of $g$ and $l_\text{s}$, therefore allowing both to be retrieved with an estimated precision of $\sim \SI{10}{\percent}$. The set of simulated curves shown in Figure~\ref{fig:grapefruit}c is obtained for $g = \num{0.7}$ and $l_\text{s} = \SI{150}{\micro\meter}$ (corresponding to an in-plane $l_\text{t} = \SI{500}{\micro\meter}$). However, quantitatively accurate figures for this particular specimen would require an analysis involving fully anisotropic simulations.

The main purpose of this experiment is to showcase the vast possibilities opened by rich time- and space-resolved outputs. Our $g$ estimation results from a collective fitting routine involving multiple spatio-temporal datasets and stands in contrast to common and less robust $g$ determination techniques involving a single scalar, such as the attenuation coefficient of a collimated beam\autocite{marchesini1989extinction, pickering1993double}. Accurate determination of single scattering anisotropy is crucial to directly probe the shape, orientation and optical properties of single microscopic scatterers and might therefore be of interest to a broad range of fields, from realistic computer rendering of participating media\autocite{pauly2000metropolis, premovze2004practical} to photo-therapeutic diagnostics of biological tissues\autocite{nilsson1998changes, mourant1998mechanisms}.
Moreover, compared to other available techniques, the determination of $g$ and $l_\text{s}$ is achieved in a single measurement session, which also simplifies the retrieval procedure. The peculiar doughnut-shaped intensity pattern that enables such a rich characterization of the scattering properties is of course not unique to this particular sample, and we observed it in a variety of different biological samples of both plant and animal origin, usually differing only by the actual contour shape of the propagating wavefront. This supports the general validity and usefulness of our proof-of-concept evaluation technique in studying extremely complex media, included biological tissues.

\paragraph{Discussion}
As illustrated, our optically gated imaging technique offers several advantages, including higher quantitative fidelity over a large field of view with respect to Fourier plane configurations and sub-ps time resolution compared to other electronically gated imaging solutions. Gaining direct access to such time scales not only enables the study of optical properties in thin or inherently minute specimens (for example, ocular fundus, vascular walls, skin dermis and dental enamel), but is fundamental to investigate light transport at the mesoscopic level. From a technical viewpoint, the main strength of our setup is its simplicity and lack of complex electronics. The inherently synchronous nature of the pulses renders the entire setup insensitive to timing jitter. Power fluctuation/drifts also do not pose a problem when measuring the mean square width evolution, which also allows us to dynamically adjust the integration time to enhance the signal-to-noise ratio. This flexibility should make it possible to replicate the setup with cheaper and more easy to use hands-off fiber lasers, therefore widening the effective applicability of the technique.

Accessing both transverse and axial transport simultaneously through a full spatio-temporal investigation allows us to correctly evaluate the applicability of the common diffusion framework, which can be subject to appreciable absorption overestimation in samples presenting a limited optical thickness or large-scale heterogeneities. An accurate retrieval of the absorption coefficient is particularly relevant in many biomedical applications where its value is directly linked to quantities such as chromophores concentration or blood oxygenation levels. The observed discrepancy, which would be not apparent with state-of-the-art spatial or temporal measurements, is revealed by comparing standard time-resolved transmitted intensity decay to the evolution of the mean square width, both of which can be acquired simultaneously. Nevertheless, the most relevant advantage offered by direct access to the transient mean square width expansion is that it enables a robust and simple interpretation. In sharp contrast to other common observables, such as the total transmittance and its decay time, the mean square width is independent of absorption and largely unaffected by refractive index contrasts and sample thickness, therefore overcoming long-standing problems posed by their precise assessment. One prominent case where this feature could be decisive is that of the experimentally observed decrease in the diffusive coefficient of a turbid slab with decreasing thickness\autocite{kop1997observation}, which was indirectly determined through the transmittance decay time. A spatio-temporal measurement in terms of the mean square width evolution would allow us to directly probe the diffusion coefficient irrespective of the exact boundary conditions, the incorrect estimation of which has been cited as a possible cause for this apparently anomalous behavior\autocite{elaloufi2002time}.

A different scenario emerges when studying samples with extremely low optical thickness, which revealed a peculiar transport regime where $l_\text{s}$ and $g$, rather than $l_\text{t}$, are the meaningful parameters determining how radiative energy spreads through the sample. This finding illustrates a further practical situation where light transport must be studied as a spatio-temporal phenomenon in its entirety. This point has not yet been properly considered when studying transport in thin samples. Moreover, we have demonstrated how this transport regime enables accurate retrieval of otherwise barely accessible transport parameters, especially for this class of samples where well-established experimental techniques are still missing.

In general, several applications will be facilitated by the three-dimensional characterization capabilities of the setup (see also SI) opening new possibilities to properly investigate structurally anisotropic media that are of paramount relevance in many industrial and biomedical applications, and whose correct theoretical modeling in terms of light transport is still an object of lively debate in the literature\autocite{PhysRevLett.98.218104, PhysRevE.89.063202}. Other relevant open questions that might be tackled directly with a spatio-temporal investigation include experimental verification of the polarization dependence predicted for the diffusion coefficient\autocite{vynck2014polarization}, as well as the occurrence of non-monotonic mean square width evolution in samples exhibiting Anderson localization\autocite{cherroret2010transverse} or the lateral expansion of L\'{e}vy type transport, where the experimental limit posed by the finite thickness of the sample might be circumvented\autocite{bertolotti2010engineering}. Similarly, an UFI measurement of intensity profiles reflected by macro-porous bulk materials could access light spreading within the first few hundred femtoseconds from which the time-dependent diffusion constant could be determined\autocite{svensson2014light} leading to information concerning the structure factor of the material, which would not be obtainable at long times.

\subsection*{Conclusions}
The field of light transport in complex or structured media still presents a multitude of experimental challenges that could significantly benefit from the advent of novel multi-domain investigation techniques. In this study, we demonstrated an experimental technique to go beyond the ps resolution barrier for spatial investigations of transient intensity profiles emerging from arbitrary samples. Gaining access to this time scale with a broad field of view is key to the characterization of a large class of samples, ranging from thin biological membranes to complex optical microdevices. As we demonstrated, spatio-temporal investigations enable the study of heterogeneous or semi-transparent media that are difficult to characterize with present state-of-the-art techniques. Further, when dealing with more standard samples, combined access to both transverse and axial transport helps reveal the presence of flaws or artifacts in the common diffusion-based data evaluation framework. Remarkably, in many practical situations, we found that low optical thickness can be exploited as an advantage rather than a drawback, therefore offering a complementary framework to study the broad class of samples that do not meet the validity conditions required by the DA. Nonetheless, the setup that we propose is generally applicable and non-invasive, usable at different wavelengths, with a broad field of view and high-time resolution. We therefore envision that it could be applied to a wide range of photonic applications for both the sub-ps physics it allows to investigate and its convenient wide-field acquisition, which does not require scanning over the region of interest.

\begin{description}[leftmargin=0cm,font=\normalfont\bfseries]\bigskip
\item[Acknowledgements] We wish to thank G. Mazzamuto, F. Utel and J. Bertolotti for technical assistance and T. Svensson, E. Alerstam, S. Gigan and P. Wasylczyk for fruitful discussions. This work is financially supported by the European Network of Excellence Nanophotonics for Energy Efficiency and the ERC through the Advanced Grant PhotBots (proj.\ ref.\ 291349).

\item[Author Contributions] M.B., L.P. and R.S. conceived and carried out the time-resolved experiments. L.P. and R.S. realized the samples, L.P. ran the simulations. All authors discussed the results and contributed to the writing of the paper.
\item[Competing Interests] The authors declare that they have no competing financial interests.
\item[Correspondence] Correspondence and requests for materials should be addressed to L.P.\\(\mail{pattelli@lens.unifi.it}).
\end{description}

\printbibliography[heading=subbibliography]
\end{refsection}

\clearpage

\beginsupplement
\begin{refsection}
\section*{Supplementary Information to: Spatio-temporal visualization of light transport in complex photonic structures}

\subsection*{Experimental setup details} The non-linear crystal used is a square $\SI{5}{\milli\meter} \times \SI{5}{\milli\meter} \times \SI{2}{\milli\meter}$ BBO crystal. Focal lengths are respectively \SI{100}{\milli\meter} for L2 and L3, and \SI{125}{\milli\meter} for the lens collecting the up-converted signal after the BBO. The final magnification obtained through the double imaging stages was roughly 1$\times$. The focusing lens L1 is chosen so to provide a point-like excitation spot with respect to the typical length scales of the sample, which in our case was set to $\sim \SI{10}{\micro\meter}$. As far as the mentioned constraints on the wavevector distribution manipulation are satisfied, a different set of optics can be employed to obtain a different magnification.
The CCD camera is a back-illuminated Andor iKon M912, with a 512$\times$512 pixel sensor.

\subsection*{Polymer samples fabrication} Investigated polymer samples are made of a commercial UV-curing acrylate optical adhesive (Norland 65) with a dispersion of rutile nanoparticles with a diameter of \SI{280}{\nano\meter} (Huntsman's Tioxide R-XL). The mixture of polymer and nanoparticles is rendered homogeneous through magnetic stirring and a ultrasonic bath ($\sim \SI{1}{\hour}$). We then let the resulting opaque paste be sucked by capillary forces inside an air gap of controlled thickness formed between two microscope glasses. The two glasses are firmly held apart by micro-spherical spacers of calibrated size. By spin coating in advance the microscope glasses with water-soluble polyvinyl alcohol we are able at a later stage to remove the glasses obtaining a flexible, free standing polymer slab. This notably allows to avoid multiple reflection of both pump and scattered light coming from the enclosing glasses.

\subsection*{Monte Carlo software}
All simulations shown have been performed with a C++ software called MCPlusPlus\autocite{MCPlusPlus} based on the standard MCML Monte Carlo code for multi-layered slab samples\autocite{wang1995mcml}. Its source code is currently under development and is freely available at \url{http://www.lens.unifi.it/quantum-nanophotonics/mcplusplus/}. Benefiting from the object-oriented paradigm of C++, the software offers a high level of abstraction, scalability, modularity and ease of maintenance. Being completely written in C++, it can be executed an any hardware and can take extensive advantage of the multi-thread capabilities of most modern CPUs for increased performance. Other features worth mentioning include python scriptability and both raw and histogrammed output in the convenient H5 binary format.

\subsection*{Probe and gate pulses characterization}
\begin{figure}
\centering
\includegraphics[width=\textwidth]{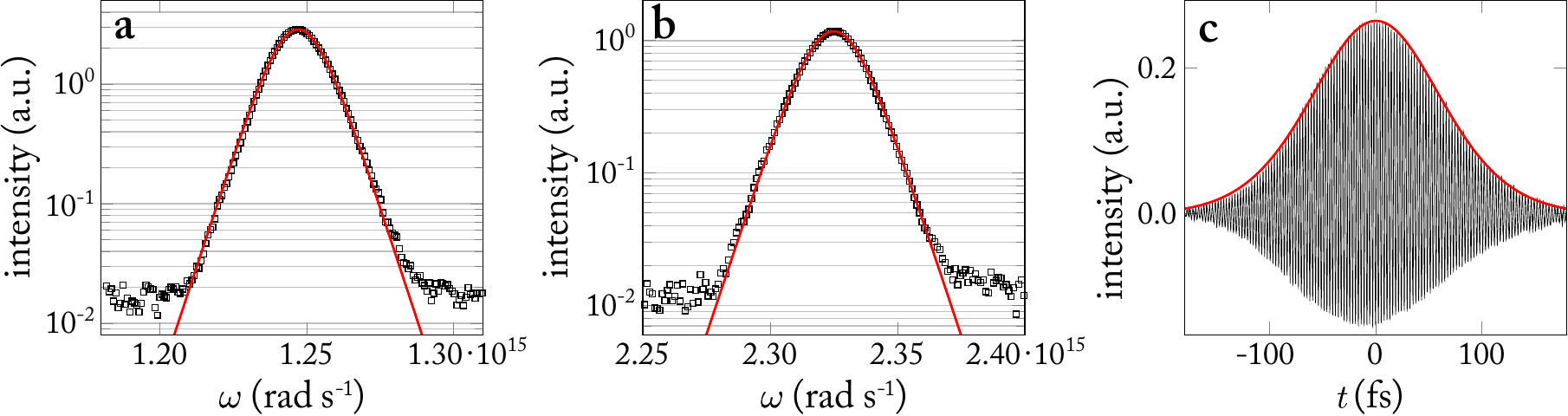}
\caption{\textbf{Pulses characterization.} \textbf{a} Spectrum of the unconverted residual (probe pulse) from the OPO, with a $\sech^2 (\omega)$ fit. \textbf{b} Spectrum of the OPO signal (gate pulse) with a $\sech^2 (\omega)$ fit. \textbf{c} Autocorrelation measurement of the residual pulse. The envelope profile fringes has been fitted with the autoconvolution of two $\sech^2 (t)$ pulses.}
\label{fig:SI_pulses}
\end{figure}

A Spectra Physics \textsc{Tsunami} Ti:Sa laser is used to pump a \textsc{Opal} optical parametric oscillator, which provides a signal at \SI{1510 +- 11}{\nano\meter}, an idler (not used in the experiment) and an unconverted residual at the same wavelength of the pump (i.e.~\SI{810 +- 5}{\nano\meter}). The average power of the signal and the residual beam is respectively \SI{190}{\milli\watt} and \SI{280}{\milli\watt}. Both pulses are characterized at the beginning of each measurement session spectrally and temporally. Figures \ref{fig:SI_pulses}a and b show typical spectra recorded for the \textsc{Tsunami} and the \textsc{Opal}, fitted with a $I(\omega) \propto \sech^2 (\frac{\pi}{2} (\omega - \omega_\text{c}) \tau_\text{p})$ function, as expected for an actively mode-locked Ti:Sa laser in a sub-\si{\pico\second} configuration (i.e.~featuring GDD compensation). Figure \ref{fig:SI_pulses}c shows an autocorrelation measurement of the residual pulse, fitted with the autoconvolution of two $\sech^2$ pulses, which can be expressed as $I_\text{ac} (t) = \gamma \, \mathrm{cosech}^2 (\gamma t) \: [\gamma \, \mathrm{cotanh} (\gamma t) - 1]$ with $\gamma = (2 \log (1+\sqrt{2}))/(\tau_\text{fwhm})$.
The full width half maximum duration of the original pulse is obtained by multiplication for the appropriate deconvolution factor, giving $\tau_\text{fwhm} = 0.6482 \times \SI{144.6}{\femto\second} = \SI{93.7}{\femto\second}$. To estimate the duration of the gate pulse we can rely on the cross-correlation measurement. Due to the lack of any analytical expression for the cross-convolution of two different $\sech^2$ pulses, we performed a fit plugging at each iteration the numerical convolution of the known pump pulse with an unknown $\sech^2$ pulse (Fig.~1b). The routine eventually returned a full width duration of \SI{134}{\femto\second} for the gate pulse. The cross-correlation was measured with and without the lenses, to make sure that the optics used for the experiment were introducing negligible dispersion. The final full width half maximum, representing the instrument response function is of approximately \SI{170}{\femto\second}. The source term in all Monte Carlo simulation showed in this paper has been modeled as a $\sech^2$ pulse of this duration, while its spatial distribution was set to a Gaussian beam with a waist of \SI{10}{\micro\meter}, as given by our focusing lens.

\subsection*{Full 3D retrieval of ballistic speed of light}
\begin{figure}
\centering
\includegraphics[width=0.8\textwidth]{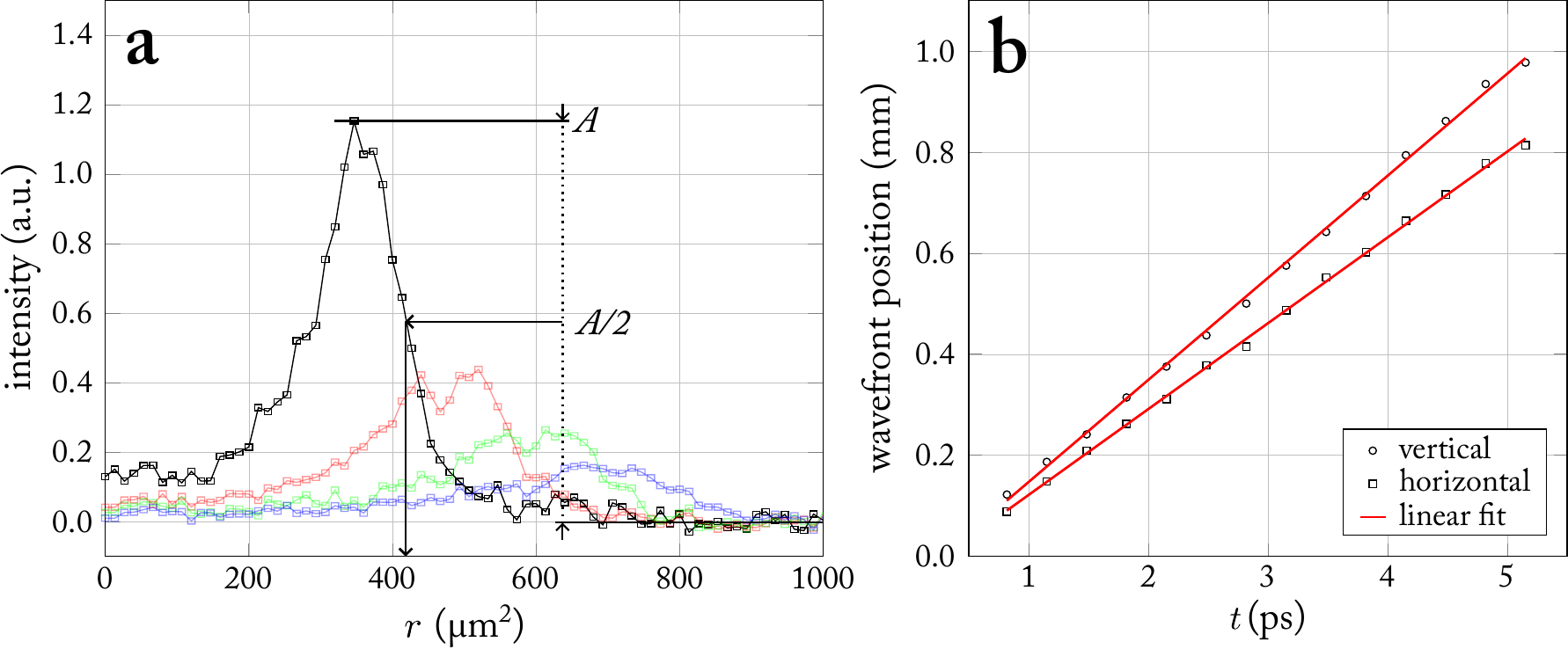}
\caption{\textbf{Grapefruit in-plane light speed retrieval.} \textbf{a} For each measured profile, the instantaneous position of its outer wavefront is calculated at half its maximum amplitude $A$ for both horizontal (shown) and vertical cross-cuts. \textbf{b} For the whole time window taken into account by our measurements the outer wavefront appears to travel ballistically with a well-defined speed. Linear fits return speeds of \SI{170}{\micro\meter\per\pico\second} ($\num{0.57}c$) and \SI{200}{\micro\meter\per\pico\second} ($\num{0.67}c$) for the horizontal and vertical axis respectively.}
\label{fig:3Dlightspeed}
\end{figure}
Spatio-temporal measurements of light traveling inside the grapefruit membrane and other analogue samples exhibit a transmitted intensity pattern which is determined by both ballistic and scattered light --- where by ``ballistic'' we refer to light that undergoes roughly just two main scattering events: one to get inside the membrane and one to be scattered out.
In particular, such ballistic component appears to persist for the whole time scale observed in our experiment and can be easily addressed by measuring the instantaneous position of the outer wavefront of the traveling pulse where, by definition, it is accumulated.
This is confirmed by the steadily linear increase of the wavefront position. The retrieved slopes directly give the light speed along a certain direction inside the slab specimen, which notably appears to be different along the $x$ and $y$ direction for the investigated sample.
Combining these results with the perpendicular delay introduced by the intervening sample whose thickness is known (cfr.\ Figure 5a) we are able to retrieve the speed of light inside the membrane along each spatial dimension.

\printbibliography[heading=subbibliography]
\end{refsection}
\end{document}